\documentclass[12pt,a4paper]{article}

\usepackage{latexsym}
\usepackage{amsmath}
\usepackage{amsfonts}
\usepackage{amssymb}

\newcommand{\qedsymb}{{\em Q.E.D.}}

\newcommand{\be}{\begin{equation}}
\newcommand{\ee}{\end{equation}}
\newcommand{\bea}{\begin{eqnarray}}
\newcommand{\eea}{\end{eqnarray}}

\renewcommand{\theequation}{\arabic{section}.\arabic{equation}}

\def\ad{{\mathrm{ad}}}                  %
\def\Ad{{\mathrm{Ad}}}                  %
\def\G{{\cal G}}                        %
\def\M{{\cal M}}                        %
\def\H{{\cal H}}                        %
\def\K{{\cal K}}                        %
\def\A{{\cal A}}                        %
\def\End{{\mathrm{End}}}                %
\def\pa{{\partial}}                     %
\def\cK{{\check {\cal K}}}              %
\def\cO{{\cal O}}                       %
\def\ri{{\mathrm{i}}}                   %
\begin{document}

\vspace*{0.5cm}
\begin{center}
{\Large \bf Spin Calogero models obtained from dynamical $r$-matrices and geodesic motion}
\end{center}

\vspace{0.2cm}

\begin{center}
L. FEH\'ER${}^{a}$ and B.G. PUSZTAI${}^b$ \\

\bigskip

${}^a$Department of Theoretical Physics, MTA  KFKI RMKI\\
1525 Budapest 114, P.O.B. 49,  Hungary, and\\
Department of Theoretical Physics, University of Szeged\\
Tisza Lajos krt 84-86, H-6720 Szeged, Hungary\\
e-mail: lfeher@rmki.kfki.hu

\bigskip
${}^b$Centre de recherches math\'ematiques, Universit\'e de Montr\'eal\\
C.P. 6128, succ. centre ville, Montr\'eal, Qu\'ebec, Canada H3C 3J7, and\\
Department of Mathematics and Statistics, Concordia University\\
7141 Sherbrooke W., Montr\'eal, Qu\'ebec, Canada H4B 1R6\\
e-mail: pusztai@CRM.UMontreal.CA

\end{center}

\vspace{0.2cm}

\begin{abstract}
We study classical integrable systems  based on the Alekseev-Meinrenken dynamical
$r$-matrices corresponding to automorphisms of self-dual Lie algebras, $\G$.
We prove that these $r$-matrices are uniquely characterized by a non-degeneracy
property and apply a construction due to Li and Xu to associate
spin Calogero type models with them.
The equation of motion of any model of this type is found to be
a projection of the natural geodesic equation
on a Lie group $G$ with Lie algebra  $\G$, and its
phase space is
interpreted as a Hamiltonian reduction of an open submanifold of the cotangent
bundle $T^*G$, using
the symmetry arising from the adjoint action of $G$ twisted
by the underlying automorphism.
This shows the integrability of
the  resulting systems and gives an algorithm to solve them.
As illustrative examples we present new models built on
the involutive diagram automorphisms of the
real split and compact simple Lie algebras, and also explain that
many further examples fit in the dynamical
$r$-matrix framework.

\end{abstract}

\newpage

\section{Introduction}
\setcounter{equation}{0}

The study of integrable many body systems initiated by Calogero \cite{Cal},
Sutherland \cite{Sut} and Moser \cite{Mos}
is popular since these systems
have interesting applications in several branches of theoretical physics
and involve fascinating mathematics.
See, for example, the reviews in \cite{Per,Nekr,CRM,SutR}.
The generalized Calogero models associated to
root systems by  Olshanetsky and Perelomov \cite{OP1}, the spin Calogero models due to
Gibbons and Hermsen \cite{GH}, and the Ruijsenaars-Schneider models \cite{RS}
are also very important.
Many approaches were applied to prove the classical integrability of  these models.
The most general among these are the construction
of a Lax representation for the equation of motion and the
realization of the system as a `projection' of some other system that
is integrable `obviously', for example the free geodesic motion on a Lie group
or on a symmetric space (see \cite{Mos, OP2, KKS} and the review \cite{Per}).
The result of Babelon and Viallet \cite{BV} linking Liouville integrability
to the form  of the Poisson brackets (PBs) of the Lax matrix is also relevant for us,
since the determination
of the $r$-matrices encoding these PBs in Calogero  models
led eventually to the questions investigated in this paper.

The $r$-matrices of the $A_n$ type spinless Calogero models found in
\cite{AT,Skly,BS} depend on the coordinates of the particles and are
not unitary. They were  re-derived by Avan, Babelon and Billey  in
\cite{ABB0} using Hamiltonian reduction of an auxiliary
non-integrable spin extension of the Calogero model. The definition
of the non-integrable model relies on a `quasi-Lax operator', whose
PBs (of the type (\ref{2.10}) below) are encoded  by certain unitary
dynamical $r$-matrices that yield the Calogero $r$-matrices upon
reduction. The unitary $r$-matrices in question were later
recognized \cite{ABB2}  to be identical to the fundamental solutions
of the so-called classical dynamical Yang-Baxter equation (CDYBE)
that appeared first in studies of the WZNW conformal field theory
\cite{BDF,Feld}. Subsequently, the geometric meaning of the CDYBE
and the classification of its solutions was investigated  by Etingof
and Varchenko \cite{EV}, and dynamical $r$-matrices have since
turned up in several contexts.

The `wonderful but mysterious' calculations of
\cite{ABB0} were understood on an abstract level
by Li and Xu, who then proposed a
method to associate an integrable
spin Calogero type
model to {\em any} spectral parameter dependent \cite{LiXu}
or independent \cite{Li2} dynamical $r$-matrix defined in \cite{EV}.
In our view
the essential  point in \cite{LiXu,Li2}
is that a `quasi-Lax operator' of a non-integrable spin
Calogero type system can be directly associated with any solution of the CDYBE,
and upon Hamiltonian reduction this yields a model admitting a Lax representation
in the usual sense.
Li and Xu also constructed new spin Calogero models by applying
their method to the rational, trigonometric and elliptic solutions
of the CDYBE  given in  \cite{EV}.
(The spin variables can  be rendered frozen  only in the $A_n$ case.)
A companion construction of
spin Ruijsenaars models was  presented  in \cite{Li1}.
These results (see also \cite{Li3,Li4}) are significant  and deserve
 further study.

The  main purpose of the present paper is to characterize the integrable spin Calogero type
models that may be associated with the remarkable  classical dynamical $r$-matrices
found by Alekseev and Meinrenken in \cite{AM}.
These trigonometric
solutions of the CDYBE are in correspondence
with  scalar product preserving automorphisms of  self-dual Lie algebras.
In particular, they contain $r$-matrices labeled by the automorphisms of the
Dynkin diagrams of the simple Lie algebras, and it seems interesting
to enquire about the corresponding integrable systems.
In our investigation we shall proceed in such a way to underline that the association
of the spin Calogero type models to the solutions of the CDYBE does not actually rely
on the machinery of Lie algebroids that features
in \cite{LiXu,Li2},
but rather it can be done in a very direct manner.
This does not mean that the algebroid-groupoid technique is not important,
since it plays a crucial role in the factorization algorithm worked out in \cite{Li2,Li1,Li4}
to solve the resulting models in the general (including the elliptic) case.
Still,  as these techniques are not widely known,
it is worth stressing that in some respects they are not essential.

In our case the algebroid-groupoid technique can also be bypassed since
it will be demonstrated that the models associated with the
Alekseev-Meinrenken $r$-matrices (\ref{3.2}) can be viewed as projections of the
natural geodesic system on a  corresponding Lie group.
We first show this at the level of the equation of motion,
which immediately yields a solution algorithm.
Then we prove that the spin Calogero phase space
obtained from the dynamical $r$-matrix construction is identical
to a reduced phase space coming from the cotangent bundle of the
corresponding Lie group,  which is reduced using the adjoint action twisted by the
underlying automorphism.
This generalizes Reshetikhin's derivation \cite{Res}
of the `principal trigonometric'  spin Calogero model by Hamiltonian reduction,
which corresponds to the identity automorphism.
Certain rational spin Calogero models will  be derived in a similar way, too,
generalizing results in \cite{Nekr,AKLM,AH,Hoch}.

As illustrative examples, we shall present new spin Calogero models
built on the non-trivial
involutive diagram automorphisms of the simple Lie algebras.
We shall also point out that many other models can be obtained
using  scalar product preserving automorphisms of reductive Lie algebras,
including for instance certain models found earlier by Blom and Langmann \cite{BL1}
and by Polychronakos \cite{Poly} by means of different methods.
The degenerate complete
integrability of the spin Calogero models corresponding to the
identity automorphism was explained in \cite{Res}  from the reduction
viewpoint (without making explicit the relationship to dynamical $r$-matrices).
It should be possible to generalize the arguments of \cite{Res} to our examples,
and one may also try to quantize
our systems by quantum Hamiltonian reduction.

In the next section, the construction of generalized
spin Calogero models based on solutions of the CDYBE is reviewed.
Subsequently, this construction is applied to those
$r$-matrices that are non-degenerate in a technical sense,
defined in Section 3, which ensures that their equation of
motion is equivalent to the Lax equation (\ref{2.17}).
This holds for the rational $r$-matrices (\ref{3.1}) and
for the Alekseev-Meinrenken $r$-matrices (\ref{3.2}).
In Section 3 the Lax equation (\ref{2.17}) is shown to be
a projection of the geodesic equation
on the underlying Lie algebra or its Lie group in correspondence with
the rational
and trigonometric cases.
Section 4 clarifies the symplectic aspect of this projection.
In Section 5 the spin Calogero Hamiltonians associated with the non-trivial
involutive diagram automorphisms of the real split  and compact simple Lie algebras
are presented.
After a brief summary, further examples and
other issues are discussed in Section 6.
In Appendix A the $r$-matrices (\ref{3.2})  are proved to be uniquely characterized by their
non-degeneracy property, which also leads to a simple description of all
quasi-triangular solutions of the CDYBE in the standard compact case.
Appendix B contains some auxiliary material.

Our main results are
Proposition 2 in Section 3 and Proposition 3 in Section 4
concerning the projection method, and the uniqueness statement of Proposition A.2
about the Alekseev-Meinrenken $r$-matrices  together with its
corollary dealing with the standard compact case.

\section{Spin Calogero  models from dynamical $r$-matrices}
\setcounter{equation}{0}

We  below  recapitulate the construction of spin Calogero type models
developed by Li and Xu  \cite{LiXu,Li2}
generalizing the earlier work of Avan, Babelon and Billey  \cite{ABB0,ABB2}.
The essential point is contained in Proposition 1 below, which
is actually very simple.
For our present purpose, we consider only spectral parameter independent
dynamical $r$-matrices defined on a self-dual {\em Abelian} subalgebra, $\K$,
of a self-dual (real or complex) Lie algebra $\G$.

The self-duality of $\G$ means that it is equipped with a
non-degenerate,
invariant, symmetric   bilinear form $\langle\ ,\ \rangle$,
which is real or complex valued depending on $\G$ being real or complex.
A subalgebra $\K\subset \G$ is called self-dual if the restriction
of  $\langle\ ,\ \rangle$ to $\K$ remains non-degenerate.
(In the physically most interesting cases $\G$ is a real Lie algebra and
 $\langle\ ,\ \rangle$  has  definite signature
on $\K$.)
The `scalar product'  $\langle\ ,\ \rangle$  gives rise to the orthogonal decomposition
\be
\G = \K + \K^\perp,
\qquad
X= X_\K + X_\perp
\qquad \forall X\in \G.
\label{2.1}\ee
A dynamical $r$-matrix associated with $\K\subset \G$ is by definition
a (smooth or holomorphic) map
\be
R: \check \K \rightarrow \End(\G)
\label{2.2}\ee
on some open domain $\check \K \subset \K$, which
is required to be antisymmetric with respect to $\langle\ ,\ \rangle$, equivariant
and subject to the CDYBE \cite{EV}.
Using the adjoint representation of $\G$, for an Abelian $\K$
the (infinitesimal)  equivariance property reads
\be
\ad_\kappa \circ R(q) = R(q) \circ \ad_\kappa
\qquad
\forall q\in \check \K,\quad \kappa \in \K.
\label{2.3}\ee
We introduce  dual bases $\{ T^i\}$ and $\{ T_j\}$ in $\K$, and
let $q^i= \langle q, T^i\rangle$ denote the corresponding coordinates on $\check \K$.
Below we use the notation
\be
\nabla_\kappa R = \kappa^i \frac{ \pa R}{\pa q^i},
\qquad
\langle X, (\nabla R) Y \rangle = T^i \langle X, \frac{ \pa R}{\pa q^i} Y \rangle
\quad \forall X,Y \in \G.
\label{2.4}\ee
The CDYBE (`with coupling $\nu$') can be formulated as the condition
\begin{equation}
E_\nu(R,X,Y)=0 \qquad \forall X, Y\in \G,
\label{2.5}\end{equation}
where $\nu$ is a constant and
\begin{eqnarray}
&& E_\nu(R,X,Y):= \frac{\nu^2}{4} [X,Y] + [ R X, R Y] -R( [X, R Y]+ [RX,Y]) \nonumber\\
&&\qquad\qquad \qquad  +\langle X, (\nabla R) Y\rangle + (\nabla_{Y_\K} R) X -
(\nabla_{X_\K} R) Y .
\label{2.6}\end{eqnarray}
In the `triangular' case $\nu=0$ and in the `quasi-triangular' case $\nu=1$.

Let us identify $\K$ and $\G$ with their dual spaces by means of the scalar
product, extend the bases $\{ T^i\}$ and $\{ T_j\}$ of $\K$
($i,j=1,\ldots, \dim(\K)$)
to dual bases $\{ T^a\}$ and
$\{ T_b\}$ of $\G$ ($a,b= 1,\ldots, \dim(\G)$), and define
$f_{ab}^c:= \langle [T_a,T_b], T^c\rangle$.
Then consider the phase space
\begin{equation}
\M:= T^* \cK \times \G^* \simeq \cK \times \K \times \G \simeq\{ (q,p, \xi)\}
\label{2.7}\end{equation}
equipped with the direct product of the natural Poisson brackets (PBs) on $T^* \cK$ and on $\G^*$.
In coordinates,
\begin{equation}
\{ q^i, p_j\}= \delta^i_j,
\qquad
\{ \xi_a, \xi_b\}= f_{ab}^c \xi_c.
\label{2.8}\end{equation}
Let us now introduce the function $L: \M \rightarrow \G$ by
\begin{equation}
L: (q,p, \xi)\mapsto p - \left(R(q)+ \frac{\nu}{2}\right)\xi.
\label{2.9}\end{equation}
One may call $L$ a quasi-Lax operator, because it satisfies PBs
of the `St Petersburg type' up  to an `anomalous term'.
This is the content of the following proposition.

\medskip
\noindent
{\bf Proposition 1 \cite{LiXu,Li2}.}
{\em
For any dynamical $r$-matrix $R$ on an Abelian, self-dual subalgebra $\K\subset \G$,
the $\G$-valued function $L$ (\ref{2.9})  on $\M$ verifies the  Poisson bracket relation
\begin{equation}
\{ L_1, L_2\} = [ R_{12}, L_1 + L_2 ] - \nabla_{\xi_\K} R_{12},
\label{2.10}\end{equation}
where $\xi_\K: (q,p,\xi)\mapsto \xi_\K$ is the evaluation map and
$R_{12}=  \langle T_a, R T_b\rangle T^a \otimes T^b$,
$L_1 = L\otimes 1$, $L_2= 1\otimes L$.
}
\medskip

Incidentally, the proof shows a stronger result.
Namely, by assuming the antisymmetry and the equivariance properties
one can calculate that
\begin{equation}
\{ L_1, L_2\} - \left([ R_{12}, L_1 + L_2 ] -
\nabla_{\xi_\K} R_{12}\right)
= \langle \xi, E_\nu(R,T^a,T^b)\rangle T_a\otimes T_b,
\label{2.11}\end{equation}
which shows that the CDYBE (\ref{2.5}) is equivalent to the PB (\ref{2.10}).

The message from (\ref{2.10}) is that
if the derivative term was absent, then the
$\G$-invariant functions of $L$ would provide a Poisson commuting family (as
usual  in classical  integrable systems).
Thus one should impose the first class constraints
\be
\xi_\K =0,
\label{2.12}\ee
and then the $\G$-invariant functions of $L$ yield a commuting
family of gauge invariant Hamiltonians.
This is a universal way whereby one obtains candidates for classical integrable systems
out of dynamical $r$-matrices.
(For particular $r$-matrices some other constraints could
kill the derivative term in (\ref{2.10}), too.)
As with constant $r$-matrices, not
all these systems are necessarily Liouville integrable, and
one may need to search for specific symplectic leaves in the reduced phase space to
obtain interesting examples.
But all these systems are at least `Lax integrable', since the PBs of $L$ with
the Hamiltonian
\be
H(q,p,\xi) = h(L(q,p,\xi))
\label{2.13}\ee
for a $\G$-invariant function $h$ on $\G$ take the following quasi-Lax form:
\be
\{ L, H\}  = [ R  V , L] - (\nabla_{\xi_\K} R)  V,
\label{2.14}\ee
where $V$ is the $\G$-valued function on the phase space defined by
\be
V(q,p,\xi) = (\nabla h)(L(q,p,\xi))
\label{2.15}\ee
with the natural $\G$-valued `gradient' of $h$.
The crucial point is that (\ref{2.14})
yields a Lax equation for $\dot{L}$ upon setting $\xi_\K=0$.
The most important Hamiltonian is
\be
h(L) = \frac{1}{2} \langle L, L \rangle,
\label{2.16}\ee
and after imposing (\ref{2.12}) it  generates
the evolution equation
\be
\dot L = [ R L, L].
\label{2.17}\ee

In the examples considered later
the operator $R(q)$ is zero on $\K$ and
$\left(R(q) + \frac{\nu}{2}\right)$
maps $\K^\perp$ to $\K^\perp$ in an invertible manner.
Assuming that $R(q)$ vanishes on $\K$,
(\ref{2.16}) gives rise  to
\be
H(q,p,\xi)  = \frac{1}{2} \langle p, p\rangle - \frac{1}{2} \langle
 \xi_\perp, R^2(q) \xi_\perp\rangle  - \frac{\nu}{2} \langle p, \xi_\K \rangle + \frac{\nu^2}{8}
 \langle \xi,\xi\rangle.
 \label{2.18}\ee
This  Hamiltonian is of the spin Calogero type since
$R(q)$ is actually either a rational
or hyperbolic function of $q$.
The fourth term in (\ref{2.18}) is a Casimir function and the third one
disappears by (\ref{2.12}).
The corresponding evolution equation can be obtained directly as
\begin{eqnarray}
\dot{\xi}_{\mathcal{K}} &=& \{\xi_\K, H\}=0,\nonumber\\
\dot{q} &=& \{q, H\}=p-\frac{\nu}{2}\xi_{\mathcal{K}}, \nonumber\\
\dot{p} &=& \{p, H\}=-\langle R(q)\xi_\perp,(\nabla R)(q)\xi_\perp\rangle, \nonumber \\
\dot{\xi}_{\perp} &=& \{\xi_\perp, H\}=[\xi,\frac{\nu}{2}p+R(q)^2\xi_\perp].
\label{2.19}\end{eqnarray}
One sees from (\ref{2.9}) that {\em if} $\left(R(q) + \frac{\nu}{2}\right)$ is invertible
on $\K^\perp$,
then the last two equations become {\em equivalent} to (2.17) upon setting $\xi_\K=0$.
We impose this constraint and are only interested in the time evolution of
the quantities that are invariant under the gauge transformations operating as
\be
(q,p,\xi_\perp, L) \mapsto (q,p, e^\kappa \xi_\perp e^{-\kappa}, e^\kappa L
e^{-\kappa}),
\label{2.20}\ee
where $\kappa$ is an arbitrary $\K$-valued function.
The transformation rule of $L$ follows from
the integrated version of (\ref{2.3}), where
in the notation we assumed that $\G$ is a matrix Lie algebra.
The time evolution of the gauge invariant
functions of $L$ is not changed if one replaces (2.17) by
\be
\dot L = [ R L -  \kappa, L]
\label{2.21}\ee
for any $\K$-valued function $\kappa$ on the (constrained) phase space.
This will be used in exhibiting the geometric meaning of the integrable systems
covered by the above formalism.

Natural analogues of Proposition 1 hold also for spectral parameter
dependent as well as for spectral parameter independent dynamical
$r$-matrices defined \cite{EV} on an arbitrary subalgebra $\K
\subset \G$. Formula (\ref{2.9}) remains essentially  the same for
all cases, the derivative $\nabla_{\xi_\K}$ in (\ref{2.10}) is
replaced by $\nabla_\chi$ with $\chi(q,p,\xi) = [q,p] + \xi_\K$ for
an arbitrary self-dual $\K$ \cite{Li2}. Direct verification is
equally easy in any case as in (\ref{2.11}). The (spectral parameter
dependent variant of the) basic formula (\ref{2.10})  first appeared
in \cite{ABB0,ABB2}, without referring to the construction of $L$
out of $R$.  The universal character of the relationship between
(\ref{2.9}), (\ref{2.10}) and the CDYBE was explained in
\cite{LiXu,Li2} as part of a general theory based on Lie algebroids.
To be more precise, \cite{LiXu} deals with spectral parameter
dependent and \cite{Li2} with spectral parameter independent
dynamical $r$-matrices. After realizing (\ref{2.10}), the
construction of integrable systems by imposing constraints arises
immediately and can be found in all the references mentioned.

\section{Interpretation as projection of geodesic motion}
\setcounter{equation}{0}

We here show that the time development of the
generalized spin Calogero model defined by the Lax equation (\ref{2.17})
can be obtained as a projection
of the free geodesic motion
on a connected Lie group $G$  with Lie algebra $\G$ if
the $r$-matrix is quasi-triangular ($\nu=1$),
or on the Lie algebra $\G$ if
the $r$-matrix is triangular ($\nu=0$).
In special cases this statement is already known \cite{Nekr,Res,AKLM},
but these papers do not explain the relationship to dynamical $r$-matrices.
Our result given by Proposition 2 below applies generally,
its  simple proof sheds light
on the geometric origin of the Lax operators (\ref{2.9})  and also
directly leads to an algorithm to solve the model.

Our assumptions on $R$ are that it is compatible with the decomposition
(\ref{2.1}), vanishes on $\K$
(which can be achieved acting by a gauge transformation on the solutions of the CDYBE)
and is {\em non-degenerate} in the sense that
the restriction of $\left(R(q) + \frac{\nu}{2} \right)$
to $\K^\perp$ is an invertible operator.
We saw that
these assumptions ensure the equivalence between (\ref{2.17})
and (\ref{2.19}) with $\xi_\K=0$.
In the triangular case one has the $r$-matrices
\be
R\colon\check\K\rightarrow\mathrm{End}(\mathcal{G}),\quad
q\mapsto R(q):=\left\{
\begin{array}{ll}
0 & \mbox{on}\:\mathcal{K},\\
\left(\mathrm{ad}_q\big|_{\mathcal{K}^{\perp}}\right)^{-1} & \mbox{on}
\:\mathcal{K}^{\perp},
\end{array}\right.
\label{3.1}
\ee
where the inverse exists on a dense domain, e.g., if $\G$ is semisimple
and $\K$ is a Cartan subalgebra.
The quasi-triangular $r$-matrices satisfying our assumptions
all have the form $R=R^\theta$ with
\be
R^{\theta}\colon\check \K\rightarrow\mathrm{End}(\G),\quad
q\mapsto R^{\theta}(q):=\left\{\begin{array}{ll}
0 & \mbox{on}\:\:\K,\\
\frac{1}{2}\left(\theta e^{\mathrm{ad}_q}\vert_{\mathcal{K}^{\perp}} +1\right)
\left(\theta e^{\mathrm{ad}_q}\vert_{\mathcal{K}^{\perp}}-1\right)^{-1} & \mbox{on}\:\:
\mathcal{K}^{\perp},
\end{array}\right.
\label{3.2} \ee
where $\theta$ is an automorphism of $\G$ that preserves also the scalar product,
$\K$ lies in the fixpoint set of $\theta$, and the inverse that occurs is
well-defined for a non-empty open subset $\check \K \subset \K$.
These quasi-triangular dynamical $r$-matrices are due to Alekseev
and Meinrenken \cite{AM} (see also \cite{ES}).
Their uniqueness under the
above mentioned conditions is a new result proved in Appendix A.
For convenience, we introduce the notation
\be
R^\theta_+(q):= R^\theta(q) + \frac{1}{2} = \left\{\begin{array}{ll}
\frac{1}{2} & \mbox{on}\:\:\mathcal{K},\\
(1-\theta^{-1} e^{-\mathrm{ad}_q}\vert_{\mathcal{K}^{\perp}} )^{-1}
& \mbox{on}\:\:\mathcal{K}^{\perp}.
\end{array}\right.
\label{3.3}\ee

To study the quasi-triangular case,
we take a connected Lie group  $G$ with Lie algebra $\G$ and suppose that
$\theta$ lifts to an automorphism $\Theta$ of $G$ (or work locally where such
a lift exists).
We restrict the considerations  to
an open submanifold $\check G\subset G$ of  `regular' elements $g$
that admit a `generalized polar decomposition':
\be
g= \Theta^{-1} (\rho) e^q \rho^{-1}
\quad\hbox{with}\quad \rho \in G, \quad q\in \check \K.
\label{3.4}\ee
A geodesic in $G$ is a curve $g(t)$ subject to the equation
\be
\frac{d}{dt} \left( g(t)^{-1} \dot{g}(t)\right)=0,
\label{3.5}\ee
where dot also denotes  `time' derivative.
We are only interested in geodesics that lie in $\check G$.

\medskip
\noindent
{\bf Proposition 2.}
{\em
Consider a curve of the product form
\be
g(t)= \Theta^{-1} (\rho(t)) e^{q(t)}\rho(t)^{-1}
\label{3.6}\ee
with smooth functions $q(t) \in \check \K$ and $\rho(t)\in G$.
Define the `spin' variable $\xi_\perp(t)$ by
\be
\xi_{\perp}(t):=R^{\theta}_+(q(t))^{-2}M_{\perp}(t)\in\mathcal{K}^{\perp}
\quad\hbox{with}\quad
M= M_\K + M_\perp := \rho^{-1} \dot{\rho}.
\label{3.7}\ee
Then the geodesic equation (\ref{3.5}) implies the same time development
for the gauge invariant functions of $q$, $\dot q$  and $\xi_\perp$ as
does the Lax equation
(\ref{2.17})  together with $p=\dot{q}$.}
\medskip

We start the proof by noticing that
\be
g^{-1}\dot{g}=\rho
\bigl(\dot{q}-(1-\theta^{-1}e^{-\mathrm{ad}_q}) M\bigr) \rho^{-1}
=\rho
\bigl(\dot{q}-(1-\theta^{-1}e^{-\mathrm{ad}_q}\vert_{\mathcal{K}^{\perp}})
M_{\perp}\bigr)\rho^{-1}.
\label{3.8}\ee
By using the definitions  in (\ref{3.3}) and (\ref{3.7}),
this is equivalently written as
\be
g^{-1}\dot{g}=\rho \bigl(\dot{q}-R^{\theta}_+(q)\xi_{\perp}\bigr)\rho^{-1}
=\rho L\rho^{-1}
\label{3.9}\ee
in terms of the Lax operator
\be
L(q, \dot q, \xi_\perp) = \dot q - R_+^\theta(q) \xi_\perp,
\label{3.10}\ee
which has the same form as (\ref{2.9}) with $\dot{q}=p$ and $\xi_\K= 0$.
Continuing with (\ref{3.9}),  (\ref{3.5}) takes the form
\be
0=\frac{\mathrm{d}}{\mathrm{d}t}(g(t)^{-1}\dot{g}(t))
=\rho (\dot{L}-[L,M])\rho^{-1}.
\label{3.11}\ee
Since $M= M_\K + R_+^\theta(q)^2 \xi_\perp$,
this in turn is equivalent to
\begin{equation}
\dot L = [L,  R_+^\theta(q)^2 \xi_\perp + M_\K].
\label{3.12}\end{equation}
The `geodesic Lax equation' (\ref{3.12}) is
to be compared with the
`spin Calogero Lax equation' (\ref{2.17}), which now reads
\be
\dot L= [R^\theta(q)L,L ] = [R_+^\theta(q)L,L]=
[L, R^\theta_+(q)^2 \xi_\perp - \frac{1}{2}\dot q].
\label{3.13}\ee
The difference of the evolutional derivatives   of $L$ defined by
(\ref{3.12}) and by (\ref{3.13})
is an infinitesimal gauge
transformation in the sense of (\ref{2.21}), and thus the proposition follows.

It is worth remarking that originally the gauge transformations (\ref{2.20})
arose from the first class constraints (\ref{2.12}), but now we can also view them
as due to the  obvious ambiguity of the parametrization of $g$ in (\ref{3.4}).
It is not difficult to show
that this parametrization  is applicable in a neighbourhood of $e^{q_0}\in G$
if and only if the $r$-matrix (\ref{3.2}) is regular at $q_0\in \K$.
In fact, for any such $q_0\in \K$ the elements of $G$ in a neighbourhood of $e^{q_0}$
can be uniquely written as\footnote{One sees this by calculating the derivative
of the map $m: \K\times \K^\perp \ni (q,\eta) \mapsto
\Theta^{-1}(e^\eta) e^q e^{-\eta}\in G$  at $(q_0,0)$.}
\be
g= \Theta^{-1}(e^{\eta}) e^q e^{-\eta},
\label{3.14}\ee
where $q$ varies in $\K$ around $q_0$ and $\eta$ varies in
$\K^\perp$ around zero.
By combining this with Proposition 2  one can in principle
solve the spin Calogero equation of motion (for gauge invariant quantities) as follows.
First, take an initial value $(q_0, p_0=\dot q_0, (\xi_\perp)_0)$.
Next, construct $L_0$ out of these data according to (\ref{3.10}).
Set $\rho_0:=e\in G$, and thus obtain from (\ref{3.5}), (\ref{3.9}) the geodesic
\be
g(t) = e^{q_0} e^{t L_0}.
\label{3.15}\ee
Then, decompose this geodesic according to (\ref{3.14}) in the form
\be
g(t)=\Theta^{-1}(\rho(t)) e^{q(t)} \rho^{-1}(t)
\quad\hbox{with}\quad
\rho(t)=e^{\eta(t)},\quad \eta(t)\in \K^\perp,
\label{3.16}\ee
which is possible at least for small values of $t$.
Finally, determine $\xi_\perp(t)$ from (\ref{3.7}) and find also $L(t)$ from (\ref{3.9}).
This procedure provides the solution of the spin Calogero equation of motion
(constrained by $\xi_\K=0$) up to a time dependent gauge transformation of the
type in (\ref{2.20}),
but this is enough since we are interested only in
gauge invariant quantities.

The non-degenerate rational $r$-matrices (\ref{3.1}) are related to geodesic
motion on the Lie algebra $\G$.
To see this let us consider a curve $X(t)\in \G$ having the product form
\be
X(t)= \rho(t) q(t) \rho(t)^{-1},
\qquad
\rho(t)\in G,\quad q(t)\in \check \K.
\label{3.17}\ee
This gives rise to $\dot X= \rho L \rho^{-1}$, where $R(q)$ is given by (\ref{3.1}) and
\be
L= \dot q - R(q) \xi_\perp
\quad\hbox{with}\quad
\xi_\perp = R(q)^{-2} M_\perp,\quad
M= M_\K + M_\perp = \rho^{-1} \dot \rho.
\label{3.18}\ee
Then the equation of motion $\frac{ d^2 X(t)} {dt^2}=0$ translates into the `geodesic Lax equation'
\be
\dot L= [L,M] = [ R(q)L - M_\K, L].
\label{3.19}\ee
This is gauge equivalent to the `spin Calogero Lax equation' (\ref{2.17}) associated
with the `principal rational $r$-matrix' (\ref{3.1}) by the construction in Section 2.

The solution algorithm sketched above must be equivalent to the one in \cite{Li2} for
the overlap of the cases considered (which actually include only the systems (\ref{5.4})).
Our derivation by the projection method was inspired by the seminal work
of Olshanetsky and Perelomov \cite{OP2} and it
appears much simpler to us than the derivation found in Section 6 of  \cite{Li2}.

For a simple
 Lie group $G$ and $\theta= \mathrm{id}$ our
Proposition 2 is consistent with Reshetikhin's
  identification \cite{Res}
  of the `principal trigonometric'
spin Calogero model (with Hamiltonian of the form in (\ref{5.4}) or (\ref{5.8})
below)
as the result  of factoring the natural Hamiltonian system on
$T^*\check G$ by the adjoint action of $G$.
Here $\check G$  is the set regular elements in the usual sense,
and $q$ (restricted to an open Weyl alcove) together
with its momentum $p$ and the spin variable in  $\K^\perp/e^\K$ provide coordinates on
the factor space $T^*\check G/G$.
Next we generalize this result to arbitrary automorphisms.

\section{Reduction of $T^*G$ by twisted conjugations}
\setcounter{equation}{0}

We have seen that the spin Calogero equation  of motion (\ref{3.13})
results from the geodesic equation on the Lie group $G$.
The geodesic motion corresponds to the Hamiltonian system $(T^*G, \Omega, H)$
with the natural symplectic form $\Omega$ and `kinetic energy' $H$ associated with
$\langle\ ,\ \rangle$.
In this section we demonstrate that the spin Calogero phase space
together with its Poisson structure and  Hamiltonian can be
identified as a Hamiltonian reduction of $(T^*G, \Omega, H)$.

As might be guessed from (\ref{3.6}) and the result of \cite{Res}
for $\Theta =\mathrm{id}$,
the reduction is by the cotangent lift of the `twisted adjoint action' of $G$ associated
with $\Theta$.
We trivialize $T^*G$ by right translations and identify $\G^*$ with $\G$ with the aid of
the scalar product.
This gives the model
\be
T^* G \simeq G \times \G = \{ (g,J)\,\vert\, g\in G,\,\,\, J\in \G\,\}
\label{4.1}\ee
with
\be
\Omega = d \langle J, (dg) g^{-1}\rangle,\qquad
H= \frac{1}{2} \langle J, J\rangle.
\label{4.2}\ee
We then consider
the $\Theta$-twisted adjoint action defined by
\be
\Ad_k^\Theta: (g,J)\mapsto (\Theta^{-1}(k) g k^{-1},  \Theta^{-1}(k) J \Theta^{-1}(k^{-1}))
\qquad
\forall k\in G,
\label{4.3}\ee
where our notation is adapted to matrix Lie algebras as usual.
The equivariant momentum map generating this Hamiltonian action of $G$,
$\psi: T^*G \rightarrow \G^*\simeq \G$,
is readily found
\be
\psi: (g,J) \mapsto \theta(J) - g^{-1} J g.
\label{4.4}\ee
Our definition in (\ref{4.3}) may look slightly cumbersome as far as the
usage of the inverse of $\Theta$ is concerned, but this is how it fits with
the form of the $r$-matrix chosen in (\ref{3.2}).

We are interested in the Marsden-Weinstein
reduced phase spaces\footnote{The factor spaces that appear here
are not everywhere smooth manifolds, but possess some
singularities.
This complication could be resolved, e.g.,
by restriction to a dense and open smooth
component
(the principal isotropy type for the group action)  or by invoking
the theory of stratified symplectic spaces and singular reduction \cite{Sjam,OR}.
In this paper we disregard these technicalities, since they do not affect
our main point.}
$(T^*G)_{\psi=\mu}/G_{\mu}$, where $\mu \in \G^*\cap \psi(T^*G)$ is chosen
arbitrarily and $G_\mu\subset G$ is its isotropy group.
We recall that $(T^*G)_{\psi=\mu}/G_{\mu}$ carries a non-degenerate Poisson structure
induced by the projection of the symplectic form $\Omega$.
These reduced phase spaces together encode the structure of $T^*G/G$, since they
correspond to  the symplectic leaves (or unions of leaves)
in $T^*G/G$ in a one-to-one manner, simply by means of projecting
$(T^*G)_{\psi=\mu}\subset T^*G$ to the $G$-orbits passing through it.
For a textbook, see e.g.~\cite{LM}.

Let $\cO$ denote the coadjoint orbit of $G$ through $-\mu$.
It will be convenient to realize $(T^*G)_{\psi=\mu}/G_{\mu}$ by using the standard
`shifting trick' as
\be
(T^*G)_{\psi=\mu}/G_{\mu} \simeq \left(T^* G \times \cO\right)_{\tilde \psi=0}/G,
\label{4.5}\ee
where $\cO$ is equipped with its own symplectic form, $\omega^\cO$,
and $\tilde\psi$ is the momentum map for the diagonal action,
$\widetilde{\Ad}_k^\Theta$,  of $G$ on $T^*G\times \cO$:
\be
\widetilde{\Ad}_k^\Theta: T^* G\times \cO \ni (g,J, \xi)\mapsto
(\Theta^{-1}(k) g k^{-1},  \Theta^{-1}(k) J \Theta^{-1}(k^{-1}), k\xi k^{-1}),
\quad \forall k\in G.
\label{4.6}\ee
The symplectic form on $T^* G \times \cO$ is given by
\be
\Omega^\cO = \Omega + \omega^\cO,
\label{4.7}\ee
and the associated momentum map reads
\be
\tilde\psi(g,J,\xi) = \theta(J) - g^{-1} J g  + \xi.
\label{4.8}\ee
The identification (\ref{4.5}) means that the
reduced symplectic structure can be induced from $\Omega^\cO$.

We  wish to show that the symplectic leaves of the spin Calogero phase space defined by
the construction of \cite{LiXu,Li2} outlined in Section 2 fill certain
dense subsets in the Marsden-Weinstein reduced phase spaces (\ref{4.5}).
The appropriate symplectic leaves result by performing the construction
of Section 2 in such a way that the spin variable is
restricted there to the orbit $\cO$ from the beginning.
To avoid notational complications,
here we make  the simplifying assumption
that $G$ is a {\em compact}, connected simple Lie group.
Denote by $G^\Theta$ the set of fixpoints of the automorphism $\Theta$ of $G$.
In the compact case
the generic orbits in $G$ under the $\Theta$-twisted adjoint action,
which we call twisted conjugacy classes,
can be parametrized
by a certain alcove in a maximal torus $K$ of $G^\Theta$ (see \cite{MW} and references therein).
It is also known \cite{Kac} that $(1- \theta^{-1} e^{-\ad_q})$ is invertible on $\K^\perp$
for generic elements $q\in \K$, where $\K$ is the Lie algebra of $K$.

\medskip
\noindent
{\bf Proposition 3.}
{\em
Consider an automorphism $\Theta$ of a compact, connected  simple Lie group $G$ and
 a Cartan subalgebra $\K$ of the Lie algebra of $G^\Theta$.
Let  $\check G \subset G$ be the $\Ad^\Theta_G$-invariant open submanifold
that consists of the $\Theta$-twisted regular conjugacy classes  represented by
$e^q$ with $q\in \check \K$, where the operator
$(1- \theta^{-1} e^{-\ad_q})\vert \K^\perp$ is invertible
and  $q\mapsto e^q$ is injective
on the open submanifold $\check \K\subset \K$.
Take an arbitrary  $\mu \in \psi(T^* \check G)$ and denote by $\cO$ the coadjoint orbit
of $G$ through $-\mu$.
Then the Marsden-Weinstein reduced phase space
$\left(T^* \check G \times \cO\right)_{\tilde\psi=0}/G$ (see (\ref{4.5}))
is identical to the phase space of the spin
Calogero system defined in Section 2 by imposing  the first class constraint
(\ref{2.12}) on $T^* \check \K \times \cO\subset \M$
(\ref{2.7}).}

\medskip

To prove Proposition 3, the key point is to solve the momentum map constraint
$\tilde\psi=0$ (\ref{4.8})  by first bringing $g\in \check G$ to its `diagonal' representative
$e^q$ with $q\in \check \K$.
After doing this, the constraint $\tilde\psi=0$
becomes
\be
\theta(J) - e^{-\ad_q} (J) + \xi =0,
\label{4.9}\ee
which is easily seen to be equivalent to
\be
\xi_\K=0
\quad\hbox{and}\quad
\theta(J) = J_\K - R_+^\theta(q) \xi_\perp,
\quad
\label{4.10}\ee
where $R_+^\theta(q)$ is defined in (\ref{3.3}).
The resulting  manifold,
$S\subset \left(T^* \check G \times \cO\right)_{\tilde\psi=0}$ given by
\be
S:= \{ (e^q, J_\K - \theta^{-1} R_+^\theta(q)\xi_\perp, \xi_\perp)\,\vert\,
q\in \check \K,\,\, J_\K \in \K,\,\, \xi_\perp \in \cO\cap \K^\perp\,\},
\label{4.11}\ee
is the gauge slice of a global partial gauge fixing for the action of $G$
on $\left(T^* \check G \times \cO\right)_{\tilde\psi=0}$.
By inspecting this partial gauge fixing one sees that
\be
\left(T^* \check G \times \cO\right)_{\tilde\psi=0} /G \simeq S/K \simeq
\check \K \times \K  \times (\cO\cap \K^\perp)/K,
\label{4.12}\ee
where $K$ is the Lie subgroup of $G^\Theta$ corresponding to $\K$ and we used
the identification
$S \simeq \check \K \times \K \times (\cO\cap \K^\perp)$ defined by the parameters
$(q, J_\K, \xi_\perp)$ on $S$.
In terms of these variables the restriction of $\Omega^\cO$ (\ref{4.7})
to $S$ takes the following form:
\be
\Omega^{\cO}\vert_S= d\langle J_\K, dq \rangle + \omega^{\cO}\vert_{\cO\cap \K^\perp}.
\label{4.13}\ee
By setting $p:= J_\K$, this becomes identical to the restriction
of the symplectic form of $T^* \check \K \times \cO$ underlying the PBs (\ref{2.8}),
which is obtained by imposing the first
class constraint (\ref{2.12}) that generate the
factorization\footnote{The singularities of $S/K$ can be avoided by restriction
to a dense open subset of ${\cal O}$ in (\ref{4.12}), like in \cite{LiXu,Li2}.}
by $K$.
This proves Proposition 3.

Notice that
upon setting $p= J_\K$,   $\theta(J)$
in the gauge $S$ (\ref{4.11}) yields precisely the constrained
quasi-Lax operator (\ref{2.9}).
It follows that the Hamiltonian also reduces as required, since
$H(g,J)=\frac{1}{2}\langle J, J\rangle  =\frac{1}{2} \langle \theta(J), \theta(J)\rangle$.

Equation (\ref{4.4}) implies that $\psi(T^*\check G)$ consists of the
coadjoint orbits of $G$ through the elements of $\K^\perp \subset \G^*$,
and it is also clear that
$(T^* \check G)_{\psi=\mu}$ is dense in
$(T^*G)_{\psi=\mu}$  for any $\mu \in \psi(T^* \check G)$.
It would be interesting to further elaborate
the relationship between the reduced systems arising from
$T^*\check G$ and from the entire $T^*G$.

For more general reductive groups,
the spin Calogero model constructed in Section 2 is essentially
(up to a discrete symmetry) identical to the factor
of $T^* \check G$ by $G$,  where $\check G$
consists of the elements $g$ in ({\ref{3.4})
and $G$ acts by twisted conjugations.
It is an open problem to explore the possible connections between
the global structure
of the reduction of $T^*G$ and the spin Calogero models
for compact as well as for non-compact groups.
In studying this problem one has to take into account the
non-conjugate Cartan subgroups of $G^\Theta$ that exist
in general and all $\Theta$-twisted conjugacy classes, whose description
is rather complicated even for $\Theta=\mathrm{id}$ \cite{Warner}.

It is easy to show  that the rational spin Calogero model
corresponding to the $r$-matrix (\ref{3.1}) is a reduction of the
geodesic system on $T^* \check \G$, where $\check \G \subset \G$
contains the adjoint orbits of $G$ through $\check \K$. For a
compact simple Lie algebra $\G$ with Cartan subalgebra $\K$ this
statement was proved in \cite{Nekr}, without mention of
$r$-matrices. See also \cite{AKLM,Hoch} for similar results about
rational spin Calogero models associated with Cartan involutions.

Let $I_h$ be an inner automorphism of $G$, operating as
$I_h: G\ni g\mapsto h g h^{-1}$ with a fixed $h\in G$, and
consider the automorphism $\tilde \Theta := \Theta \circ I_h$.
The corresponding twisted adjoint actions of $G$ on $T^*G$
are related by the formula
\be
\Ad_k^{\tilde \Theta} = {\cal L}_{h^{-1}} \circ \Ad_k^\Theta \circ {\cal L}_h
\qquad
\forall k\in G.
\label{4.14}\ee
Here ${\cal L}_h: T^*G \rightarrow T^*G$ is the symplectomorphism
${\cal L}_h: (g,J) \mapsto (hg, hJ h^{-1})$, which also preserves the kinetic energy.
It follows from (\ref{4.14}) that the symmetry reductions of the geodesic system on
$T^* G$ by the $\Theta$-twisted and the $\tilde \Theta$-twisted adjoint actions
lead to  reduced Hamiltonian systems that are isomorphic by
a map induced from ${\cal L}_h$.
By Proposition 3, this implies the isomorphic nature of the integrable
spin Calogero models that can be obtained by applying the method of \cite{LiXu,Li2}
to `input data' related by an inner automorphism of $\G$.
In fact,
one expects on general grounds that
only different cosets in $\mathrm{Aut}(\G)/ \mathrm{Int}(\G)$
correspond to  significantly different integrable systems,
and
this could be confirmed
directly as well without using the Hamiltonian reduction of $T^*G$.

\section{Examples based on involutive diagram automorphisms
of real split and compact simple Lie algebras}
\setcounter{equation}{0}

As an illustration,  we  here present
the spin Calogero systems associated
with a real split or a compact simple Lie algebra equipped with
an involutive automorphism induced from the Dynkin diagram.
The systems corresponding to the
non-trivial diagram automorphisms of the $A_n$, $D_n$ and $E_6$ algebras
have not been described before, while those associated with the
trivial diagram automorphism appear also, e.g., in \cite{LiXu,Li2,Li3,Res}.
To keep the presentation short, in all cases we write down only the
$r$-matrix and the Hamiltonian after imposing the constraint $\xi_\K=0$.
See eqs.~(5.4), (\ref{5.8}) and (5.14), (\ref{5.17}) for the Hamiltonians
corresponding to the trivial and non-trivial automorphisms, respectively.

First, we introduce some notations relying on standard results about Lie algebras.
Let $\A$ be a simple complex Lie algebra with Killing form
$\langle\ ,\ \rangle$, Cartan subalgebra $\H$, set of simple roots $\Pi=\{ \varphi_k\}$ and
positive roots $\Phi_+$.
Choose root vectors  $X_{\pm \varphi_k}$ normalized so that
$\langle X_{\varphi_k}, X_{-\varphi_k}\rangle =1$.
Any symmetry $\tau$  of the Dynkin diagram of $\A$ extends to an automorphism
of $\Phi_+$ as well as to an automorphism of $\A$ by the requirement
$\tau(X_{\pm \varphi_k})= X_{\pm \tau(\varphi_k)}$ for any $\varphi_k\in \Pi$.
The diagram automorphism $\tau$ of $\A$ commutes with the Chevalley
automorphism $\sigma$ defined by $\sigma(X_{\pm \varphi_k}) = - X_{\mp \varphi_k}$.
We fix root vectors $X_{\pm \varphi}$ in such a way that
\be
X_{-\varphi} = - \sigma(X_\varphi)
\quad\hbox{and}\quad
\langle X_\varphi, X_{-\varphi}\rangle = 1
\qquad
\forall \varphi \in \Phi_+.
\label{5.1}\ee
We denote by $\H_r$ the real span of the elements
$T_{\varphi_k}:= [X_{\varphi_k}, X_{-\varphi_k}]$.

With the above conventions,
the split (normal) real form of $\A$  is the  Lie algebra $\A_r$ defined  by
the real span of the base elements  $T_{\varphi_k}$ and $X_{\pm \varphi}$.
To describe the example provided by the data
$\G:= \A_r$, $\K:= \H_r$, $\theta=\mathrm{id}$, we expand the `spin'
variable $\xi\in \A_r$ as
\be
\xi= \xi_{\H_r}+ \sum_{\varphi \in \Phi}
\xi_{\varphi} X_\varphi
\quad
\hbox{with}\quad
\xi_{\H_r}\in \H_r.
\label{5.2}\ee
Equation (\ref{3.2}) gives the standard $r$-matrix
\be
R(q) \xi = \frac{1}{2} \sum_{\varphi \in \Phi} \xi_\varphi
\coth \frac{\varphi(q)}{2} X_\varphi\,,
\label{5.3}\ee
where $q\in \check \H_r$ is such that this function is well-defined.
Upon imposing the constraint $\xi_{\H_r}=0$,
the spin Calogero Hamiltonian (\ref{2.18})  takes the form
\be
H(q,p,\xi) = \frac{1}{2} \langle p, p \rangle
-\frac{1}{4} \sum_{\varphi \in \Phi_+} \frac{\xi_\varphi
\xi_{-\varphi} }{ \sinh^2 \frac{\varphi(q)}{2}}.
\label{5.4}\ee

The compact real form $\A_c$ of $\A$ is the real span of the base elements
$\ri T_{\varphi_k}$ together with
\be
Y_\varphi:= \frac{\ri}{\sqrt{2}} (X_\varphi + X_{-\varphi})
\quad\hbox{and}\quad
Z_\varphi:= \frac{1}{\sqrt{2}} (X_\varphi - X_{-\varphi})
\quad
\forall \varphi \in \Phi_+.
\label{5.5}\ee
For $\G:=\A_c$, $\K:=\ri \H_r$ and $\theta=\mathrm{id}$ using the expansion
\be
\xi= \ri \xi_{\H_r} + \sum_{\varphi \in \Phi_+} \left( \eta_\varphi Y_\varphi +
\zeta_\varphi Z_\varphi \right),
\label{5.6}\ee
we have
\be
R(\ri q) \xi= \frac{1}{2} \sum_{\varphi \in \Phi_+} \cot \frac{\varphi(q)}{2}
\left(\eta_\varphi Z_\varphi - \zeta_\varphi Y_\varphi\right).
\label{5.7}\ee
In this case the spin Calogero variables are
$(\ri q, \ri p, \xi)\in \check \K \times \K \times \K^\perp$,
and the Hamiltonian  reads
\be
H(\ri q,\ri p,\xi) = -\frac{1}{2} \langle p, p \rangle
-\frac{1}{8} \sum_{\varphi \in \Phi_+} \frac{\eta_\varphi^2 +
\zeta_{\varphi}^2 }{ \sin^2 \frac{\varphi(q)}{2}}\,,
\label{5.8}\ee
where $q\in \H_r$ is such that $\sin \frac{\varphi(q)}{2}\neq 0$.

Now let us consider an involutive automorphism $\tau$ of $\A$
induced from a Dynkin diagram automorphism as outlined above.
Notice that $\tau$ preserves the split and the compact real forms as
well as the real Cartan subalgebra $\H_r$.
Denote by $\A^\pm$, $\A_r^\pm$, $\A_c^\pm$, $\H^\pm$ and $\H_r^\pm$ the
corresponding eigensubspaces of $\tau$ with eigenvalues $\pm 1$.
It is known that $\A^+$ is a complex simple Lie algebra with Cartan subalgebra $\H^+$,
and $\A^-$ is an irreducible module of $\A^+$ whose
non-zero weights have multiplicity one \cite{Kac}.
Furthermore, $(\H_r^+, \A_r^+)$ and $(\ri \H_r^+, \A_c^+)$ are the split
and compact real forms of $(\H^+, \A^+)$ with the associated real irreducible modules
$\A_r^-$ and $\A_c^-$, respectively.
Convenient bases of all these spaces can be obtained from the above considered
(`re-normalized') Chevalley
basis of $\A$ by the  `folding' procedure.
We recall this in Appendix B, and next summarize the result only.

Let $\Delta$ be the set of  roots of $(\H^+, \A^+)$
with associated root vectors $X^+_{\alpha}$ for $\alpha\in \Delta$.
Let $\Gamma$ denote the set of non-zero  weights for $(\H^+, \A^-)$
with corresponding weight vectors $X^-_{\lambda}$ for $\lambda \in \Gamma$.
We can choose these elements so that
\be
\langle X^+_\alpha, X^+_{-\alpha}\rangle  = 1 = \langle X^-_\lambda, X^-_{-\lambda}\rangle
\quad\hbox{and}\quad
\sigma(X^+_\alpha)= - X^+_{-\alpha},
\quad
\sigma(X^-_\lambda)=- X^-_{-\lambda}.
\label{5.9}\ee
The folding procedure ensures that these bases are such that
$\A_r^+$ is the real span of $\H_r^+$ together with the $X^+_{\alpha}$ for
$\alpha\in \Delta$, and $\A_r^-$ is spanned by $\H_r^-$ and the $X_\lambda^-$
for $\lambda \in \Gamma$.
Let $\Delta_+$ and $\Gamma_+$ denote positive roots and weights.
The folding procedure also ensures that
$\A_c^+$ is the real span of $\ri \H_r^+$ together with
basis vectors
\be
Y^+_\alpha:= \frac{\ri}{\sqrt{2}} (X^+_\alpha + X^+_{-\alpha})
\quad\hbox{and}\quad
Z^+_\alpha:= \frac{1}{\sqrt{2}} (X^+_\alpha - X^+_{-\alpha})
\quad
\forall \alpha \in \Delta_+,
\label{5.10}\ee
and
$\A_c^-$ is the real span of $\ri \H_r^-$ together with
basis vectors
\be
Y^-_\lambda:= \frac{\ri}{\sqrt{2}} (X^-_\lambda + X^-_{-\lambda})
\quad\hbox{and}\quad
Z^-_\lambda:= \frac{1}{\sqrt{2}} (X^-_\lambda - X^-_{-\lambda})
\quad
\forall \lambda \in \Gamma_+.
\label{5.11}\ee

Now we consider the system associated with the data
$\G:= \A_r$, $\K:= \H_r^+$, $\theta:= \tau$.
We write $\xi\in \A_r$ as
\be
\xi = \xi_{\H_r}^+ + \xi_{\H_r}^- +
\sum_{\alpha \in \Delta} \xi^+_\alpha X^+_\alpha
+
\sum_{\lambda \in \Gamma} \xi^-_\lambda X^-_\lambda
\quad\hbox{with}\quad
\xi_{\H_r}^\pm \in \H_r^\pm.
\label{5.12}\ee
According to (\ref{3.2})
\be
R^\tau(q) \xi = \frac{1}{2} \sum_{\alpha \in \Phi} \xi_\alpha^+
\coth \frac{\alpha(q)}{2} X^+_\alpha
+ \frac{1}{2} \sum_{\lambda \in \Gamma} \xi_\lambda^-
\tanh \frac{\lambda(q)}{2} X^-_\lambda\,,
\label{5.13}\ee
where this function is smooth for $q\in \check \H_r^+$.
After imposing the constraint $\xi_{\H_r}^+=0$, we obtain from (\ref{2.18})
the Hamiltonian
\be
H(q,p,\xi) = \frac{1}{2} \langle p, p \rangle
-\frac{1}{4} \sum_{\alpha \in \Delta_+}
\frac{\xi_\alpha^+ \xi_{-\alpha}^+ }{ \sinh^2 \frac{\alpha(q)}{2}}
+ \frac{1}{8} \langle \xi_{\H_r}^-, \xi_{\H_r}^-\rangle
+\frac{1}{4} \sum_{\lambda \in \Gamma_+}
\frac{\xi_\lambda^- \xi_{-\lambda}^- }{ \cosh^2 \frac{\lambda(q)}{2}}.
\label{5.14}\ee

In the corresponding compact case,
with data $\G:= \A_c$, $\K:= \ri \H_r^+$, $\theta=\tau$, we parametrize
$\xi \in \A_c$ as
\be
\xi=\ri \xi_{\H_r}^+ +  \ri \xi_{\H_r}^-
+ \sum_{\alpha \in \Delta_+} \left( \eta^+_\alpha Y_\alpha^+ +
\zeta^+_\alpha Z^+_\alpha \right) +
 \sum_{\lambda \in \Gamma_+} \left( \eta^-_\lambda Y_\lambda^- +
\zeta^-_\lambda Z^-_\lambda \right).
\label{5.15}\ee
For $\ri q\in \check \K \subset \ri \H_r^+$, we then obtain
\be
R^\tau(\ri q) \xi= \frac{1}{2} \sum_{\alpha \in \Delta_+} \cot \frac{\alpha(q)}{2}
\left(\eta^+_\alpha Z^+_\alpha - \zeta^+_\alpha Y^+_\alpha\right)
-
\frac{1}{2} \sum_{\lambda \in \Gamma_+} \tan \frac{\lambda(q)}{2}
\left(\eta^-_\lambda Z^-_\lambda - \zeta^-_\lambda Y^-_\lambda\right).
\label{5.16}\ee
After imposing the first class constraints $\xi_{\H_r}^+=0$,
the spin Calogero variables are
$(\ri q, \ri p, \xi)\in \check \K \times \K \times \K^\perp$,
and the Hamiltonian  reads
\be
H(\ri q,\ri p,\xi) = -\frac{1}{2} \langle p, p \rangle
-\frac{1}{8} \sum_{\alpha \in \Delta_+} \frac{ (\eta^+_\alpha)^2+
(\zeta^+_{\alpha})^2 }{ \sin^2 \frac{\alpha(q)}{2}}
- \frac{1}{8} \langle \xi_{\H_r}^-, \xi_{\H_r}^-\rangle
+\frac{1}{8} \sum_{\lambda \in \Gamma_+} \frac{ (\eta^-_\lambda)^2+
(\zeta^-_{\lambda})^2 }{ \cos^2 \frac{\lambda(q)}{2}}.
\label{5.17}\ee

We finish by listing the positive roots $\Delta_+$ and weights $\Gamma_+$
that arise for the involutive diagram automorphisms of the classical
Lie algebras.

If $\A= D_{n+1}$, then $\A^+= B_n$ and the module $\A^-$ is isomorphic to the
defining representation of $B_n$.  Therefore
\be
\Delta_+= \{ e_k \pm e_l,\, e_m\,\vert\, 1\leq k < l \leq n,\,\, 1\leq m\leq n\,\},
\quad
\Gamma_+ = \{ e_m \,\vert\, 1\leq m\leq n\,\}.
\label{5.18}\ee
One may realize $\H_r^+$ as the space of real diagonal matrixes, $q$, of the form
\be
q=\mathrm{diag}(q_1,\ldots, q_n, 0, 0, - q_n, \dots, -q_1)
\quad\hbox{and}\quad
e_m: q \mapsto q_m.
\label{5.19}\ee

If $\A=A_{2n-1}$, then $\A^+=C_n$ and
\be
\Delta_+= \{ e_k \pm e_l,\, 2e_m\,\vert\, 1\leq k < l \leq n,\,\, 1\leq m\leq n\,\},
\quad
\Gamma_+ = \{ e_k \pm e_l \,\vert\, 1\leq k<l\leq n\,\}.
\label{5.20}\ee
Now $\H_r^+$ can be realized as the space of real diagonal matrixes written as
\be
q=\mathrm{diag}(q_1,\ldots, q_n, - q_n, \dots, -q_1)
\quad\hbox{and}\quad
e_m: q \mapsto q_m.
\label{5.21}\ee

For $\A=A_{2n}$ one has $\A^+=B_n$ with $\Delta_+$ as in (\ref{5.18}) and
\be
\Gamma_+= \{ e_k \pm e_l,\, e_m,\, 2e_m\, \vert\, 1\leq k < l \leq n,\,\, 1\leq m\leq n\,\}.
\ee
Now $\H_r^+$ consists of the real diagonal matrices
\be
q=\mathrm{diag}(q_1,\ldots, q_n,0, - q_n, \dots, -q_1)
\quad\hbox{and}\quad
e_m: q \mapsto q_m.
\label{5.22}\ee

In the $\A=E_6$ case $\A^+ = F_4$ and the weights of its representation on $\A^-$
can be found in \cite{Kac}.
Of course, one can also construct an integrable spin Calogero model by using the
third order diagram automorphism of $D_4$.

If in the above we allow all variables  to be complex, then the models given by
(\ref{5.4}) and (\ref{5.8}) (or respectively by (\ref{5.14}) and (\ref{5.17}))
yield the holomorphic spin Calogero model belonging to the complex Lie algebra
$\A$ equipped with the corresponding automorphism.
In other words, we obtained different real forms of complex spin Calogero models
using the real forms $\A_r$ and $\A_c$ of $\A$.
For general considerations about real forms of complexified Hamiltonian systems
with the spinless Calogero models as examples, we refer to \cite{GKMV}.

\section{Discussion}
\setcounter{equation}{0}

One of the original motivations behind the theory of constant and dynamical classical
$r$-matrices is the potential application to build integrable systems using the $r$-matrices.
The aim of the present paper
has been to contribute to this theory by characterizing the systems
that correspond by the construction of \cite{LiXu,Li2} to the non-degenerate
dynamical $r$-matrices (\ref{3.2}).
(See also the uniqueness statement of Proposition A.2 in Appendix A.)
We demonstrated that these spin Calogero type models
are projections of the natural geodesic system on the underlying Lie group $G$,
where the automorphism enters through the twisted adjoint action of $G$.
The Hamiltonian reduction picture that we developed permits to
understand the exact solvability of the reduced systems, in principle from any point of
view ranging from the proof of
complete integrability to the explicit construction of the solutions.

We presented  new integrable models built on the diagram
automorphisms of the real split and compact simple Lie algebras.
To illustrate the considerable range  of applicability of
the dynamical $r$-matrix construction,  we now  mention further
examples that could be developed utilizing $r$-matrices of the form (\ref{3.2}).
For instance, one could apply the construction
to the Cartan involutions of the simple real Lie algebras.
Cartan involutions are singled out since their use guarantees
the definite character of the kinetic energy of the corresponding models.
New models are expected to arise from those Cartan
involutions that are outer automorphisms of non-split real Lie algebras.
In fact, this holds for the real forms
$su^*(2n)$ and  $so(2p+1, 2q+1)$ ($p\neq q$) of the classical complex Lie algebras
and for some real forms of $E_6$.
A rich  set of other examples can  be derived  by considering
direct sum Lie algebras composed of $N>1$ identical copies of
 a self-dual Lie algebra $\mathcal{A}$.
 The scalar product of $\A$ induces a scalar product on
\begin{equation}
\mathcal{G}:=\mathcal{A}\oplus\mathcal{A}\oplus\cdots\oplus\mathcal{A}
\label{7.1}
\end{equation}
in such a way that one obtains a scalar product preserving automorphism $\theta$ of $\G$ by
the formula
\begin{equation}
\theta\colon\mathcal{G}\rightarrow\mathcal{G},\quad
\theta(u_1,u_2,\ldots,u_N):=(\tau u_N,\tau u_1,\ldots,\tau u_{N-1}),
\label{7.2}
\end{equation}
where $\tau$ is a scalar product preserving automorphism of $\A$.
In the simplest case $\tau=\mathrm{id}$ and $\theta$ is
the cyclic permutation automorphism.
We have inspected this case by taking an arbitrary  simple Lie
algebra for $\A$ and taking $\K$ as the diagonal embedding of a Cartan subalgebra of
$\A$ into $\G$.
Then it turned out that the
dynamical $r$-matrix construction reproduces (for $\A=A_n$) certain generalized spin Calogero
models found earlier in \cite{BL1,Poly} by different methods.
These examples  and their generalizations for affine Lie algebras will be further
studied elsewhere.

A very interesting  question that remains to be investigated is whether all
spin Calogero type models  that may be associated with
arbitrary solutions of CDYBE  (\ref{2.5}) and its spectral parameter dependent version
can be understood as Hamiltonian reductions of suitable
`obviously integrable' systems.
The quantization of the spin Calogero models is another
important problem that appears to be largely open.

Finally, let us discuss the generalization of the construction of Section 2
to dynamical $r$-matrices based on {\em non-Abelian} subalgebras,
which were introduced in \cite{EV} and appeared afterwards in various
contexts (see e.g.~\cite{Feh,EE} and references therein).
In fact,
the main examples of such $r$-matrices are the
`extended versions' of the
 Alekseev-Meinrenken $r$-matrices (\ref{3.2}) given
on the full fixpoint set, say $\G_0 \subset \G$, of  the automorphism $\theta$
by the formula \cite{AM}
\be
R_{\mathrm{ext}}^{\theta}\colon\check \G_0\rightarrow\mathrm{End}(\G),\quad
q\mapsto R_{\mathrm{ext}}^{\theta}(q):=\left\{\begin{array}{ll}
f(\ad_q) & \mbox{on}\:\:\G_0\\
\frac{1}{2}\left(\theta e^{\mathrm{ad}_q}\vert_{\G_0^\perp} +1\right)
\left(\theta e^{\mathrm{ad}_q}\vert_{\G_0^\perp}-1\right)^{-1} & \mbox{on}\:\:
\G_0^{\perp}
\end{array}\right.
\label{7.3}\ee
with the analytic function $f(z) = \frac{1}{2}\coth \frac{z}{2}-\frac{1}{z}$.
Now,
even if $\G_0$ is non-Abelian, one can still use (\ref{2.9}) to define a quasi-Lax operator
$L(q,p,\xi)$ on $T^* \check \G_0\times \G^*\simeq \check \G_0\times \G_0\times \G$,
but then the anomalous term in  (\ref{2.10}) becomes
$\nabla_\chi R_{\mathrm{ext}}^\theta$ with $\chi(q,p,\xi) = [q,p] + \xi_{\G_0}$.
This statement can be found in \cite{Li2} and is also readily
verified similarly to Proposition 1.
Therefore, in the non-Abelian case one has to impose the constraint $\chi=0$
to get rid of the anomalous term.
If a non-Abelian $\G_0$ contains an appropriate Abelian $\K$,
then $\chi=0$ represents a larger set of constraints than
$\xi_\K=0$ does, and $\chi=0$ also generates a larger gauge group.
In the end,
further supposing for simplicity that $\G$ is {\em compact},
it is possible to show that the reduced phase space of the
`non-Abelian construction' defined by imposing the first class constraint
$\chi=0$ on $T^*\check\G_0 \times \G^*$
is {\em the same} as the one
obtained from the corresponding
`Abelian construction' that operates by imposing $\xi_\K=0$ on $T^*\check\K \times \G^*$.
(If $\G$ is reductive, then $\K$ must be a Cartan subalgebra of $\G_0$
for $R^\theta$ (\ref{3.2}) to be well-defined.
A slight complication for a non-compact  real  reductive $\G$ arises since
$\G_0$ admits non-conjugate Cartan subalgebras in general.)
By the argument just sketched, we do not expect the
$r$-matrices based on non-Abelian subalgebras to lead to interesting new integrable models.
 This observation motivated us in the first place to restrict ourselves
throughout this paper to dynamical $r$-matrices based on Abelian subalgebras.
Our argument is closely related to the mapping between different $r$-matrices defined by
means of Dirac
reduction of the space of dynamical variables as studied in \cite{FGP}.
 The connection between the Abelian and non-Abelian variants
of the dynamical $r$-matrix construction of spin Calogero type models
is elaborated in more detail in \cite{Varna}

\renewcommand{\theequation}{\arabic{section}.\arabic{equation}}
\renewcommand{\thesection}{\Alph{section}}
\setcounter{section}{0}

\section{On the non-degenerate Abelian dynamical $r$-matrices}
\setcounter{equation}{0}
\renewcommand{\theequation}{A.\arabic{equation}}

In this appendix we
show that the $r$-matrices (\ref{3.2}) found in \cite{AM}
can be uniquely characterized among the
solutions of the CDYBE (with coupling $\nu=1$)
by requiring the additional properties (\ref{T7}) and (\ref{T8}) that ensure
the equivalence of the Lax equation (\ref{2.17}) with the
Hamilton equation (\ref{2.19}) upon imposing the constraint $\xi_\K=0$ and setting $p=\dot{q}$.
The constrained Hamilton equation always implies the Lax equation, but is not
necessarily equivalent to it in general.
For compact Lie algebras the non-degeneracy condition (\ref{T7}) holds automatically,
which allows us to obtain a rather strong result about the solutions of the CDYBE
in this case.

Before presenting our results we introduce
a convenient notion of Cayley transformation.
Let $R \in \End(\G)$ be an antisymmetric linear operator on a self-dual Lie algebra $\G$
such that the operators $R_\pm:= R \pm \frac{1}{2}\mathrm{id}_\G$
 are \emph{invertible}.
 (This hold for all antisymmetric operators if $\G$ is a real Lie algebra
 with  `scalar product' $\langle\ ,\ \rangle$ of definite signature.)
We call
\be
C :=R_+ R_-^{-1}
\label{T1}\ee
the Cayley transform of $R$.
We then also have the inverse Cayley transformation formula
\be
R =\frac{1}{2}(C+1)(C-1)^{-1}.
\label{T2}
\ee
Note that $C\in \End(\G)$ is an `orthogonal' operator , i.e.,
\be
\langle C X,CY\rangle=\langle X,Y\rangle,\quad
\forall X,Y\in\G.
\label{T3}\ee
In fact, (\ref{T1}) and (\ref{T2}) represent  a bijective correspondence
between the `antisymmetric' operators $R$ for which $R_\pm$ are invertible and
the `orthogonal' operators $C$ that do not admit the eigenvalue $+1$.
(Up to some signs, this becomes the usual Cayley transformation for $\G=so(3)$.)

Let $\mathcal{K}\subset \G$ be an arbitrary (not necessarily Abelian)
self-dual subalgebra with a non-empty open, connected domain $\check \K\subset \K$.
Assuming that
\be
R\colon\check{\mathcal{K}}\rightarrow\End(\G),\quad
q\mapsto R(q)
\label{T4}\ee
is antisymmetric and  $R_\pm(q)$
 are \emph{invertible}
for all $q\in\check{\mathcal{K}}$, we introduce the Cayley transform
$C\colon\check{\mathcal{K}}\rightarrow\mathrm{End}(\mathcal{G})$ by pointwise
application of (\ref{T1}).
One can then verify

\medskip
\noindent
\textbf{Lemma A.1.}
\emph{The map $R$ is
$\mathcal{K}$-equivariant, iff its Cayley transform $C$ is $\K$-equivariant,
i.e.,
\be
\left(\nabla_{[T,q]}R\right)(q)=\left[\mathrm{ad}_T,R(q)\right]
\quad\Leftrightarrow \quad
\left(\nabla_{[T,q]}C\right)(q)=\left[\mathrm{ad}_T,C(q)\right]
\quad
(\forall T\in\mathcal{K},\:\forall q\in\check{\mathcal{K}}).
\label{T5}
\ee
$R$ solves the CDYBE (\ref{2.5}) with $\nu=1$, iff $C$ satisfies the equation
\begin{eqnarray}
C(q)[X,Y]-[C(q)X,C(q)Y]
+(C(q)-1)\langle C(q)X,(\nabla C)(q)Y\rangle\nonumber\\
-(\nabla_{(C(q)Y-Y)_{\mathcal{K}}}C)(q)X
+(\nabla_{(C(q)X-X)_{\mathcal{K}}}C)(q)Y=0
\label{T6}\end{eqnarray}
for all $q\in\check{\mathcal{K}}$ and $X,Y\in\mathcal{G}$.
}

Here, we are interested in antisymmetric $r$-matrices,
$R\colon\mathcal{\check\K}\rightarrow\End(\G)$,
defined on an  \emph{Abelian} self-dual subalgebra $\mathcal{K}\subset \G$
that are {\em compatible} with the decomposition
$\G=\K+ \K^\perp$
and satisfy the following {\em non-degeneracy} condition:
\be
R_-(q)\vert_{\K^\perp} \:\:
\hbox{is invertible $\forall q\in\check{\mathcal{K}}$.}
\label{T7}
\ee
The compatibility really means that $R(q) \K \subset \K$ since then
$R(q) \K^\perp \subset \K^\perp$ follows from the antisymmetry.
For simplicity,
we  assume that the mapping $R: \check \K\to \End(\G)$ is smooth
or holomorphic for real or complex $\G$, respectively.
Without further loss of generality, we can impose the auxiliary condition
\be
R(q)\vert_{\K} =0,\quad  \forall q\in\check{\mathcal{K}}.
\label{T8}
\ee
This can be achieved, because (as is well known and easy to check)
the restriction of  the operator $R(q)$ to $\K$
is required by the CDYBE to yield an {\em arbitrary} closed $2$-form, $F$, on $\check \K$  by
\be
F(q) = \sum_{ij} \langle T_i, R(q) T_j\rangle dq^i \wedge dq^j
\label{T9}\ee
with a basis $T_i$ of $\K$ and corresponding coordinates $q^i$ of $q\in \check \K$.
If (\ref{T7}) and (\ref{T8}) hold, then we can use the Cayley transform (\ref{T1})
to prove the following result.

\medskip
\noindent
\textbf{Proposition A.2.}
\emph{
The (smooth or holomorphic)  $\K$-equivariant solutions of the
CDYBE (\ref{2.5}) with $\nu=1$, on an Abelian
self-dual subalgebra $\K\subset \G$, that are non-degenerate in the sense (\ref{T7})
and satisfy also (\ref{T8}) are precisely the maps of the form
\be
R(q)X=\frac{1}{2}(\theta e^{\mathrm{ad}_q}+1)
\left((\theta e^{\mathrm{ad}_q} -1)\vert_{\K^\perp}\right)^{-1} X,\quad
\forall X\in \K^\perp,\, q\in\check{\mathcal{K}},
\label{T10}
\ee
where $\theta$ is a scalar product preserving automorphism of $\G$ for which
$\theta X=X$ for any $X\in \K$ and the inverse in (\ref{T10}) exists.
}

\medskip
\noindent
\textbf{Proof.}
By the Cayley transformation formula (\ref{T1}) the assumption
$R(q)|_{\mathcal{K}}=0$ translates into the equivalent equation
$C(q)|_{\mathcal{K}}=-\mathrm{id}_{\mathcal{K}}$, and
our aim is to determine $C(q)|_{\mathcal{K}^{\perp}}$.
To do that, first notice that in our case the $\mathcal{K}$-equivariance
property (\ref{T5}) reads
\be
[\mathrm{ad}_T,C(q)]=0,\quad
\forall q\in\check{\mathcal{K}},\forall T\in\mathcal{K}.
\label{T11}
\ee
Now, let us turn our attention to the equation (\ref{T6})
for $C$. When $X,Y\in\mathcal{K}$, then (\ref{T6}) becomes an obvious
identity and gives no information. Next,
consider the case when
$X\in\mathcal{K}$ and $Y\in\mathcal{K}^{\perp}$. Then we easily get
\be
(\nabla_X C)(q)Y=[X,C(q)Y].
\label{T12}
\ee
We get the same equation when $X\in\mathcal{K}^{\perp}$
and $Y\in\mathcal{K}$, up to change of letters.
The most interesting case occurs when $X,Y\in\mathcal{K}^{\perp}$.
Then, using the above two relations one finds that
(\ref{T6}) can be written as
\be
C [X,Y]-[C X,C Y]+2[C X,C Y]_{\mathcal{K}}=0,
\label{T13}
\ee
which is in fact equivalent to the following system of equations
\be
[C X,C Y]_{\mathcal{K}} = [X,Y]_{\mathcal{K}},\quad
\left[C X,C Y\right]_{\perp} =
C \left[X,Y\right]_{\perp}.
\label{T14}
\ee
Now, let us introduce the function
\be
\theta_\perp\colon\check{\mathcal{K}}\rightarrow\mathrm{GL}(\mathcal{K}^{\perp}),
\quad
q\mapsto\theta_\perp(q)
:=\left(C(q)e^{-\mathrm{ad}_q}\right)\vert_{\mathcal{K}^{\perp}}.
\label{T15}
\ee
Equation  (\ref{T12}) with (\ref{T11}) implies that
this function satisfies the differential equation
\be
(\nabla_X\theta_\perp)(q)=0,\quad
\forall q\in\check{\mathcal{K}},\forall X\in\mathcal{K}.
\label{T16}
\ee
Hence, $\theta_\perp(q)$ is necessarily constant,
$\theta_\perp(q)=\theta_\perp\:\:(\forall q\in\check{\mathcal{K}})$.
We extend this constant operator onto the whole
Lie algebra $\mathcal{G}$ by the formula
\be
\theta\colon\mathcal{G}\rightarrow\mathcal{G},\quad
\theta:=\left\{
\begin{array}{ll}
\mathrm{id}_{\mathcal{K}} & \mbox{on}\:\:\mathcal{K},\\
\theta_\perp & \mbox{on}\:\:\mathcal{K}^{\perp}.
\end{array}\right.
\label{T17}
\end{equation}
Then, one readily  checks that $C(q)$ satisfies equations
(\ref{T3}), (\ref{T11}), (\ref{T12})
and (\ref{T14}) if and only if
the extended operator $\theta$ (\ref{T17}) is a
scalar product preserving automorphism of the self-dual
Lie algebra $\mathcal{G}$. We have just demonstrated  that under our
conditions the  Cayley transform (\ref{T1}) of $R(q)$ has the form
\be
C(q)=\left\{
\begin{array}{ll}
-\mathrm{id}_{\mathcal{K}} & \mbox{on}\:\:\mathcal{K},\\
\theta e^{\mathrm{ad}_q} & \mbox{on}\:\:\mathcal{K}^{\perp},
\end{array}\right.
\label{T18}
\ee
where $\theta\in\mathrm{Aut}(\mathcal{G})$ is a
scalar product preserving automorphism with the property
$\theta\vert_{\mathcal{K}}=\mathrm{id}_{\mathcal{K}}$. So, by
applying the inverse Cayley transformation (\ref{T2})
we obtain the Proposition.
\qedsymb
\medskip

Now we wish to point out some consequences. First note that if the
self-dual, Abelian subalgebra $\K\subset \G$ is {\em maximal} as
subalgebra, then $R(q) \K \subset \K$ follows from the equivariance
property. Thus in this case the $\K^\perp$-component of any
non-degenerate solution of the CDYBE (with coupling $\nu=1$) has the
form (\ref{T10}), while its $\K$-component is furnished by a closed
$2$-form $F$ according (\ref{T9}). This gives a characterization of
the non-degenerate dynamical $r$-matrices on the Cartan subalgebras
of the (real or complex) simple Lie algebras. If $\K$ is a Cartan
subalgebra of a {\em complex} simple $\G$, then all automorphisms
that fix $\K$ pointwise have the form $\theta=e^{\ad_v}$ for a
constant $v\in \K$, and Proposition A.2 reproduces a result of
Etingof and Varchenko (see Theorem 3.1 in \cite{EV}). More
interestingly, notice that Proposition A.2 can be applied to the
classification of the $r$-matrices on  Abelian subalgebras of {\em
compact} Lie algebras, since in this case $R_\pm $ are automatically
invertible for any antisymmetric $R$. Hence we obtain the following
`dynamical analogue' of the well-known uniqueness result \cite{Soib}
on constant $r$-matrices of compact Lie algebras.

\medskip
\noindent {\bf Corollary A.3.}
{\em
The $\K$-equivariant solutions of the CDYBE (\ref{2.5}) with $\nu=1$ on a
Cartan subalgebra $\K$ of a compact simple Lie algebra $\G$ are precisely the maps
$q \mapsto R(q)$ for which $R(q)\vert_{\K}$ corresponds to
a  closed $2$-form  by (\ref{T9})  and
\be
R(q)\vert_{\K^\perp} = \frac{1}{2} \coth \bigl(\frac{1}{2}
\ad_{q +v}\vert_{\K^\perp}\bigr)
\label{T19}\ee
with an arbitrary constant $v\in\K$.
The smoothness of $R$ requires that $\varphi(q+v)\notin 2\pi\ri \mathbb{Z}$
for any root $\varphi$ with respect to (the complexification of) $\K$.}

\medskip
To obtain the corollary we used that the automorphisms of a compact simple  $\G$
whose fixpoint sets contain the Cartan subalgebra  $\K$
are of the form $e^{\ad_v}$ with some $v\in \K$.
It can also be checked that from all (non-degenerate or not) solutions of the CDYBE that
exist in the corresponding complex case \cite{EV}
precisely those giving (\ref{T19}) survive the restriction to the compact real form.


For clarity, let us finally remind that in the CDYBE (\ref{2.5}) the
constant $\nu$ can be replaced by $\beta^2\nu$ if one replaces
$R(q)$ by $\beta R(\beta q)$ with a constant $\beta\neq 0$.
If $\nu \neq 0$, then for a complex $\G$ the value of
$\nu$ can thus be scaled to $1$, while for a real $\G$ the two different
possibilities $\nu=1$ and $\nu=\ri$ arise in principle.
We have chosen $\nu=1$ since this is required for Proposition 1.
(The same choice is needed for applications in the WZNW model \cite{Feh}.)
Up to gauge transformations,  for a simple $\G$  all non-constant
solutions with $\nu\neq 0$ that are known to us belong to $\nu=1$.
However, for a compact $\G$ the constant solutions have $\nu=\ri$,
and to our knowledge it has not been investigated if
$\nu=\ri$ is excluded for non-constant solutions of the CDYBE or not.

\section{Folding by an involutive  diagram automorphism}
\setcounter{equation}{0}
\renewcommand{\theequation}{B.\arabic{equation}}

For convenience, we here summarize the `folding procedure' referred to in Section 5.

We denote by $\tau$ the non-trivial involutive automorphism of the Dynkin diagram of
$\A\in \{ A_n, D_n, E_6\}$ as well as its extension to $\A$ and to the root system $\Phi$.
Let us decompose the set of positive roots in the form
\be
\Phi_+ = \Xi\cup \Psi \cup \tau(\Psi),
\label{A.1}\ee
where the elements of $\Xi$ are fixed by $\tau$, and the 2-point orbits of
$\tau$ in $\Phi_+$ are represented by $\Psi$.
Since $\tau$ is a Cartan preserving involution of $\A$,
any choice of root vectors enjoys the property
$\tau(X_\varphi) = c_\varphi X_{\tau (\varphi)}$ with some constants $c_\varphi$ subject to
$c_\varphi c_{\tau (\varphi)}=1$.
If the root vectors are chosen as described in Section 5, then
these constants satisfy also the relations
$c_{-\varphi} = c_\varphi$ and $c_\varphi c_{-\varphi}=1$.
The first relation follows since $\tau$ commutes with the Chevalley
automorphism  $\sigma$, and the second one is implied by
$1= \langle X_\varphi, X_{-\varphi} \rangle = \langle \tau(X_\varphi), \tau(X_{-\varphi})\rangle$.
As a consequence, $c_\varphi^2 =1$ and $c_\varphi = c_{\tau(\varphi)}$
for any $\varphi \in \Phi$.
We can decompose $\Xi$ as  $\Xi=\Xi^+\cup \Xi^-$ with
\be
\Xi^\pm := \{ \varphi \in \Phi_+\,\vert\, \tau(\varphi)=\varphi,\, c_\varphi = \pm 1\,\}.
\label{A.2}\ee
For any $\varphi \in \Phi$ denote by $\bar \varphi$ the restriction of
$\varphi$ to $\H^+$.
It is known (and is easy to check) that $\bar \varphi \neq 0$
$\forall\varphi\in \Phi$ and different $\tau$-orbits in $\Phi$ yield different
functionals on $\H^+$ upon restriction.

In fact, a system of positive roots of $(\H^+, \A^+)$,  introduced in Section 5,
is provided by
\be
\Delta_+ = \{ \bar \varphi \, \vert\, \varphi \in \Psi\cup \Xi^+ \,\},
\label{A.3}\ee
and the positive weights of $(\H^+, \A^-)$ are furnished by
\be
\Gamma_+ = \{ \bar \varphi \, \vert\, \varphi \in \Psi\cup \Xi^- \,\}.
\label{A.4}\ee
The root vectors used in Section 5 can be taken to be
\be
X_{\pm \bar\varphi}^+ = \frac{1}{\sqrt 2} \left(X_{\pm \varphi} + \tau(X_{\pm \varphi})\right)
\quad \forall \varphi \in \Psi,
\qquad
X_{\pm \bar \varphi}^+ = X_{\pm\varphi} \quad \forall \varphi \in \Xi^+,
\label{A.5}\ee
and the weight vectors can be  chosen as
\be
X_{\pm \bar\varphi}^- = \frac{1}{\sqrt 2} \left(X_{\pm \varphi} - \tau(X_{\pm \varphi})\right)
\quad \forall \varphi \in \Psi,
\qquad
X_{\pm \bar \varphi}^- = X_{\pm\varphi} \quad \forall \varphi \in \Xi^-.
\label{A.6}\ee
It is easy to see that the generators provided by this `folding procedure' have
the properties claimed in Section 5.

In the most complicated example  $\A= A_{2n}$, we have
$\Phi_+= \{ \epsilon_k - \epsilon_l \,\vert \, 1\leq k < l\leq (2n+1)\,\}$.
With the usual conventions
$\tau: \epsilon_k \mapsto - \epsilon_{2n+2 -k}$ ($\forall k=1,\ldots, 2n+1$).
Then we obtain
\be
\Psi = \{\, \epsilon_k - \epsilon_l, \epsilon_k - \epsilon_{2n+2-l}, \epsilon_m - \epsilon_{n+1}\,
\vert\, 1\leq k < l \leq n,\, 1 \leq m\leq n\,\},
\label{A.7}\ee
\be
\Xi= \Xi^- =\{ \epsilon_m - \epsilon_{2n+2-m}\,\vert\, 1\leq m\leq n\,\}.
\label{A.8}\ee
The restriction of the above elements of $\Psi$ and $\Xi^-$ to $\H^+$ gives
\be
\bar\epsilon_k - \bar\epsilon_l = e_k - e_l,\quad
\bar \epsilon_k - \bar \epsilon_{2n+2-l}= e_k + e_l,
\quad
\bar\epsilon_m - \bar\epsilon_{n+1}=e_m,
\quad
\bar\epsilon_m - \bar\epsilon_{2n+2-m}=2e_m.
\label{A.9}\ee
In the usual realization
$A_{2n}= sl_{2n+1}$,  $\tau: E_{a,b} \mapsto (-1)^{b-a+1} E_{2n+2-b, 2n+2-a}$
with the elementary matrices $E_{a,b}$ for $a,b=1,\ldots, 2n+1$.
Then one readily verifies (\ref{A.9}) by using that
$\epsilon_a$ maps $Q= \mathrm{diag}(Q_1,\ldots, Q_{2n+1})\in \H$ to
$Q_a$ for $a=1,\ldots, 2n+1$
and $e_k$ maps $q=\mathrm{diag}(q_1,\ldots, q_n, 0, -q_n,\ldots, -q_1)\in \H^+$ to $q_k$
for $k=1,\ldots, n$.
The claimed properties of the root and weight vectors can also be checked on this model,
but, of course, all the calculations can be done in a model independent manner as well.

\bigskip
\bigskip
\noindent{\bf Acknowledgements.}
The work of L.F. was supported in part by the Hungarian
Scientific Research Fund (OTKA) under the grants
T043159, T049495,  M045596  and by the EU networks `EUCLID'
(contract number HPRN-CT-2002-00325) and `ENIGMA'
(contract number MRTN-CT-2004-5652).
B.G.P. is grateful for support by a CRM-Concordia Postdoctoral
Fellowship and he especially  wishes to thank J. Harnad for
hospitality in Montreal.
We are also indebted to J. Balog and I. Marshall for useful comments on the manuscript.

\end{document}